\documentclass[aps,prb,twocolumn,floatfix,floats,showpacs,superscriptaddress,raggedbottom,amsmath,amssymb]{revtex4}
\pdfoutput=1

\usepackage[
		colorlinks=true, %Turns of the ugly box around and links and makes them colored instead
		pdfauthor={Kristinn Torfason, Chi-Shung Tang and Vidar Gudmundsson},
		pdftitle={Coherent magnetotransport and time-dependent transport through split-gated quantum constrictions}]{hyperref}

\usepackage{graphicx,xcolor,latexsym}
\usepackage{dcolumn}
\usepackage{amsmath}

\def\b#1{\mathbf{#1}}

\def\nn{\nonumber \\}

\def\eq#1{Eq.\ (\ref{#1})}
\def\fig#1{Fig.\ \ref{#1}}

\newcommand{\ud}{\mathrm{d}}

\begin{document}

%\preprint{}

\title{Coherent magnetotransport and time-dependent transport\\through split-gated quantum constrictions}

\author{Kristinn Torfason}
\affiliation{Science Institute, University of Iceland, Dunhaga 3,
        IS-107 Reykjavik, Iceland}

\author{Chi-Shung Tang}
\email{cstang@nuu.edu.tw}
\affiliation{Department of Mechanical
Engineering, National United University, 1, Lienda, Miaoli 36003,
Taiwan}

\author{Vidar Gudmundsson}
\email{vidar@raunvis.hi.is}
\affiliation{Science Institute, University
of Iceland, Dunhaga 3, IS-107 Reykjavik, Iceland}

\date{\today}

\begin{abstract}
The authors report on modeling of transport spectroscopy in split-gate controlled
quantum constrictions.  A mixed momentum-coordinate representation is
employed to solve a set of time-dependent Lippmann-Schwinger equations
with intricate coupling between the subbands and the sidebands. Our
numerical results show that the transport properties are tunable by
adjusting the ac-biased split-gates and the applied perpendicular
magnetic field. We illustrate time-modulated quasibound-state
features involving inter-sideband transitions and the Aharonov-Bohm
oscillation characteristics in the split-gated systems.
\end{abstract}


 % 72.10.-d Theory of electronic transport; scattering mechanisms
 % 73.23.-b Electronic transport in mesoscopic systems
 % 73.63.Nm Quantum wires
 % 73.63.Kv Quantum dots
 % 72.40.+w Photoconduction and photovoltaic effects
 % 74.50.+r Tunneling phenomena; point contacts, weak links, Josephson effects
 % 75.47.-m Magnetotransport phenomena; materials for magnetotransport
 % 75.75.+a Magnetic properties of nanostructures

%\keywords{}

\maketitle

\section{Introduction}

Coherent transport phenomena in mesoscale conductors with various
geometries have attracted much attention over recent years due to their
potential in the investigation of various resonance or bound-state
features,\cite{wees1988,Liu1996,Li2000,Clerk2001,Rokhinson2006} imaging
coherent electron wave
flow,~\cite{Topinka2000,Kim2003,Mendoza2005,Hackens2006} and electrical
switching
effects.~\cite{Son2005,Hartmann2007,Valsson2008,Morfonios2009} The
conductance $G$ is a fundamental property of quasi-one-dimensional
systems close to twice the quantum unit of conductance $G_0 = 2e^2/h$,
where $-e$ denotes the charge of an electron, the factor of 2 accounts
for spin degeneracy, and $h$ is Planck's constant.  Moreover, the
conductance depends sensitively on the particular arrangement of
scatterers as well as the applied external fields in the mesoscopic
system.

Conducting structures subject to magnetic fields or periodically
varying voltages are essential fundamental entities in mesoscopic
physics. The momenta of electrons are allowed to undergo a robust
change if a magnetic field is applied, which in turn dynamically
modifies the transport properties of a quantum system.  It was
indicated that the conductance involving Aharonov-Bohm (AB)
interference\cite{Aharonov1959} as a function of magnetic field
exhibits step-like structures.\cite{Taylor1992,Camino2005} Recently,
differential conductance of an AB interferometer was measured as a
function of the bias voltage.\cite{Sigrist2007}  Varying either the
magnetic field or the electrostatic confining potentials allows the
interference to be tuned.

A mesoscale system driven by an external time-dependent potential
allows charge carriers to make coherent inelastic
scattering\cite{Tang1996,Chung2004} involving inter-subband and
inter-sideband transitions.\cite{Chu1996,Tang2003}  A number of
interesting time-dependent transport related issues have been
investigated such as time-modulated quasibound-state (QBS)
features,~\cite{Tang1999,Wu2006} quantum pumping
effects,~\cite{Thouless1983,Switkes1999,Tang2001,Moskalets2002,%
Agarwal2007,Stefanucci2008} and nanomechanical
rectifiers.~\cite{Mal'shukov2005,Kaun2005,Pistolesi2006}  If the
driving frequency is comparable to the subband level spacing, the
pumping becomes nonadiabatic and manifests reversely shifted partial
gap in the transmission as a function of energy.\cite{Tang2001} The driving
force behind nonadiabatic pumping is the coherent inelastic multiple
backscattering involving either absorption or emission of a quantized
photon energy.  It is noteworthy that these effects are applicable to
design a tunable current source or reversely to create a quantum motor
driven by the generated electric current,\cite{Qi2009} or fast
manipulations for quantum information processing.\cite{Braun2008}

Of particular interest are investigations of the interplay between the
various effects of electron transport in mesoscopic systems. In this
study the main stress falls on the investigation of tunable quantum
magnetoconductance that could be manipulated by the ac-biased quantum
point contacts (QPCs) in the presence of a magnetic field perpendicular
to the two-dimensional electron gas (2DEG) plane. This can be
achievable by designing a specific size and geometry of the quantum
constriction by controlling split-gate voltages for manipulating the
coupling strength between the leads and the open cavity region confined
by the QPCs.

In Sec.\ II, we specify the setup and the geometry of the
two-dimensional quantum channel structure, and present the theoretical
framework as well as computational approach. In Sec.\ III, the main
transport spectroscopy features are demonstrated and discussed along
with the underlying dynamical mechanisms.  Concluding remarks are
presented in Sec.\ IV.

\section{Model and time-dependent Lippmann-Schwinger approach}

The system under investigation is composed of split-gates confined QPCs
embedded in a quantum channel with parabolic confining potential
$V_{\rm c}(y) = \frac{1}{2} m^*\Omega_0^2 y^2$, and hence the electrons
are transported through the broad channel with characteristic energy
scale $\hbar\Omega_0$ in the transverse direction.  The electrons
incident from the reservoirs propagating in the $\b{x}$-direction
impinge on the QPC system scattered by a local time-periodic potential
$V_{\rm sc}(x,y,t)$ under the influence of a perpendicular magnetic
field.  The system is supposed to be fabricated from a modulation doped
GaAs/AlGaAs heterostructure hosting a 2DEG system. The Hamiltonian thus
consists of
\begin{eqnarray}
 {\cal H}(t) &=&  -\frac{\hbar^2}{2m^*} \left(\nabla^2 -
 \frac{2i}{l^2} y \partial_x - \frac{y^2}{l^4} \right) \nn
 &&+  V_{\rm c}(y) + V_{\rm sc}(x,y,t)   \, ,
\end{eqnarray}
where we choose a value of effective mass $m^*$=$0.067m_e$ of the
charge carriers to corresponding to a GaAs-based 2DEG and $l$=$\hbar/(e B)$
denotes the magnetic length of an electron.   The local time-dependent
scattering potential
  \begin{equation}
    V_{\rm sc}(x,y,t) = V_{s}(x,y) + \sum_i V_{t;i}(x,y)\cos(\omega t + \phi_{i})
  \end{equation}
contains a static part with spatial dependent strength $V_s$ and a
time-dependent part with spatial dependent strength $V_t$, driving
frequency $\omega$ and phase $\phi$.

In the presence of a magnetic field ${\bf B}$=$B{\hat{\bf z}}$, the
time-dependent Schr\"{o}dinger equation ${\cal H}(t)\Psi = i\hbar\partial_t
\Psi$ is inseparable in the ($x,y$)-coordinates, but is separable in
the mixed momentum-coordinate
representation,~\cite{Gurvitz1995,Gudmundsson2005} namely transforming
the total wave function $\Psi(x,y,t)$ into the wave function
$\Psi(p,y,t)$ and expanded in terms of the eigenfunctions $\phi_n(y,p)$
of an ideal quantum channel
\begin{equation} \label{eq:wavefunc-expan}
    \Psi(p,y,t) = \sum_n \phi_n(y,p) \psi_n(p,t) \, .
\end{equation}
Due to the effects of the Lorentz force, the eigenfunctions of the parabolic
confinement $\phi_n(y,p)$ are shifted by $y_0 = p \omega_c/(\beta^2
\Omega_{\omega})$ with $\beta = \sqrt{m^*\Omega_\omega/\hbar}$ being
the reciprocal of the effective magnetic length of the system.  Here
the effective confining strength $\hbar\Omega_\omega =
\hbar\sqrt{\omega_c^2 + \Omega_0^2}$ under a magnetic field is related to
the bulk cyclotron frequency $\omega_{\rm c}=eB/(m^*c)$ and the
characteristic frequency $\Omega_0$ for the parabolic confinement.  The
resulting equation after the expansion is a coupled nonlocal integral
equation in the momentum space describing the electron propagation of
an asymptotic state occupying subband $n$ along the $\b{x}$-direction
that can be expressed as
\begin{eqnarray}\label{psi-pt}
 i\hbar\partial_t \psi_n(p,t) &=& E_n(p) \psi_n(p,t) \nonumber\\
 &&+ \sum_{n^{\prime}} \int \frac{\ud q}{2\pi} V_{n,n^{\prime}}(p,q,t)\psi_{n^{\prime}}(q,t)\, .
\end{eqnarray}
Here the electron energy $E_n(p) =  E_n(0) + K(p)$ of the subband $n$
contains the subband threshold $E_n(0) = \left(n +
\frac{1}{2}\right)\hbar\Omega_\omega$ for conduction ($n$ = $0$, $1$,
$\ldots$) that is determined by the lateral confinement and the
effective kinetic energy
\begin{equation}
 K(p) =  \frac{(\hbar\Omega_0)^2}{(\hbar\Omega_\omega)^2} \frac{\hbar^2 p^2}{2m^*} \, .
\end{equation}
In the integrand of \eq{psi-pt}, the overlap integral
\begin{equation}\label{eq:matrix-elements}
   V_{n,n^\prime}(p,q,t) = \int \ud y \ud x e^{-i(p-q)x} \phi_n^*(y,p)V(x,y,t)\phi_{n^\prime}(y,q)
\end{equation}
constructing the matrix elements of the scattering potential indicates
the electrons in the subband $n$ making inter-subband transitions to
the intermediate states $n^\prime$.

Due to the periodicity in time of the driving field, the time-dependent
wave function with incident energy $E_0$ and the driving potential
can be transformed into the frequency domain, namely
\begin{equation}
\label{t-exp-psi}
    \psi_n(p,t) = \sum_{m = -\infty}^{\infty} e^{-iE_m t/\hbar} \psi_n^m(p)
\end{equation}
and
\begin{equation}
\label{t-exp-V}
    V_{n,n^\prime}(p,q,t) = \sum_{m^\prime = -\infty}^{\infty} e^{-im^{\prime}\omega t} V_{n n^\prime}^{m^\prime}(p,q)\, ,
\end{equation}
where the quasi-energy $E_m = E_0 +m\hbar\omega$ with $m$ and
$m^\prime$ indicating the indices of sidebands induced by the external
driving field.  The magnitude of the wave vector $k_n^m$ along the
$\b{x}$-direction in the $(n,m)$ intermediate state can be expressed as
\begin{equation}
\frac{1}{2}\left( \frac{k_n^m}{\beta} \right)^2
\frac{(\hbar\Omega_0)^2}{\hbar\Omega_\omega} = E_m  - E_n(0)\, .
\end{equation}
This is convenient for us to obtain the multiple scattering identity
containing intricate inter-subband and inter-sideband transitions
\begin{eqnarray}\label{eq:def-green}
 &&\left\lbrace \left(\frac{k_n^m}{\beta} \right)^2 - \left(\frac{q}{\beta} \right)^2 \right\rbrace
 \psi_n^m(q) \nonumber \\
 &=& \sum_{m^{\prime} n^{\prime}} \int \frac{\ud p}{2\pi} \, \widehat{V}_{n,n^{\prime}}^{m-m^{\prime}}(q,p) \psi_{n^{\prime}}^{m^{\prime}} (p) \, ,
\end{eqnarray}
  where we have defined
\begin{equation}
    \widehat{V}_{n,n^{\prime}}^{m-m^{\prime}} (q,p) = 2 \frac{(\hbar \Omega_{\omega})^2}{(\hbar \Omega_0)^2}
                                           \frac{\beta}{\hbar \Omega_{\omega}} V_{n,n^{\prime}}^{m-m^{\prime}} (q,p)
\end{equation}
for simplicity.  From \eq{eq:def-green}, we can define the Green
function of the $(n,m)$ state as
\begin{equation} \label{eq:green_func}
    \left\lbrace \left(\frac{k_n^m}{\beta} \right)^2 - \left( \frac{q}{\beta} \right)^2 \right\rbrace G_n^m(q) = 1
\end{equation}
and the corresponding incident wave obeys
  \begin{equation} \label{eq:in_wave_sol}
    \left\lbrace \left(\frac{k_n^m}{\beta} \right)^2 - \left( \frac{q}{\beta} \right)^2 \right\rbrace \psi_n^{m,0}(q) = 0 \, .
  \end{equation}
After some algebra, we can obtain the Lippmann-Schwinger equation for
the Fourier components of the wave function scattered into the $(n,m)$
state
\begin{eqnarray} \label{eq:lipp_schw_wavefunc}
\psi_n^m (q) &=& \psi_n^{m,0} (q) + G_n^m (q) \nonumber \\
    &&\times \sum_{r,m^{\prime}} \int \frac{\ud (p/\beta)}{2\pi}\,
            \widehat{V}_{n,r}^{m-m^{\prime}} (q,p) \psi_r^{m^{\prime}} (p)
  \end{eqnarray}
by taking all the intermediate $(r,m^{\prime})$ states into account.
We note that the above equation is not suitable for numerical
calculations because the incident wave $\psi_n^{m,0}(q)$ is a delta
function in the Fourier space.  To achieve exact numerical calculation,
one has to define the $T$ matrix
 \begin{eqnarray}\label{eq:T-matrix_full}
    T_{n^\prime,n}^{m^\prime,m}(q,p) &=& V_{n^\prime,n}^{m^\prime - m} \nn
                                     &+& \sum_{r,s} \int \frac{\ud k}{2\pi} V_{n^\prime,r}^{m^\prime - s}(q,k) G_r^s(k)
                                     T_{r,n}^{s,m}(k,p) \nn
 \end{eqnarray}
that couples all the intermediate $(n,m)$ and $(n^\prime,m^\prime)$
states.  The scattering potential is expanded in the Fourier series in
\eq{t-exp-V} and the matrix elements calculated according to
\eq{eq:matrix-elements}. This yields a connection between the sidebands
in the system which can be seen clearly when the matrix elements for
the potential are inserted into \eq{eq:T-matrix_full} for the $T$
matrix,
\begin{eqnarray}\label{eq:T-matrix_eq}
    &&T_{n^{\prime},n}^{m^{\prime},m} (q,p) =  V_{s,n^{\prime} n}(q,p) \delta_{m^{\prime}-m,0} \nn
    &&+ \frac{1}{2} V_{t, n^{\prime} n} (q,p) (\delta_{m^{\prime}-m,-1} + \delta_{m^{\prime}-m,1}) \nn
    &&+      \sum_r \int \frac{\ud k}{2\pi} V_{s,n^{\prime} r} (q,k) G_r^{m^{\prime}} (k) T_{r,n}^{m^{\prime},m} (k,p) \nn
    &&+      \frac{1}{2} \sum_r \int \frac{\ud k}{2\pi} V_{t, n^{\prime} r}^{+} (q,k) G_r^{m^{\prime}+1} (k) T_{r,n}^{(m^{\prime}+1),m} (k,p) \nn
    &&+      \frac{1}{2} \sum_r \int \frac{\ud k}{2\pi} V_{t, n^{\prime} r}^{-} (q,k) G_r^{m^{\prime}-1} (k) T_{r,n}^{(m^{\prime}-1),m} (k,p) \nn
\end{eqnarray}
where
\begin{equation}
  V_{t; n^\prime r}^{\pm}(q,k) = \sum_i V_{t\,i;\, n^\prime r}(q,k) e^{\pm i \phi_i}\,
\end{equation}
from which we can see that adjacent sidebands are coupled. This allows
us to write the momentum space wave function in terms of the $T$ matrix
  \begin{eqnarray}\label{eq:wave-func-t}
    \psi_{n^\prime}^{m^\prime}(q) &=& \psi_{n^\prime}^{m^\prime,0}(q) + G_{n^\prime}^{m^\prime}(q)\nonumber\\
                                  &&\times \sum_{rs} \int \frac{\ud k}{2\pi} T_{n^\prime,r}^{m^\prime,s}(q,k)\psi_r^{s,0}(k)\, .
  \end{eqnarray}
The transmission coefficients or amplitudes can be found by
constructing the full wave function by inserting
Eq.\ \eqref{eq:wave-func-t} back into the expansions done previously,
i.e.\ Eqs.\ \eqref{eq:wavefunc-expan} and\ \eqref{t-exp-psi}. After
performing the inverse Fourier transform into the coordinate
representation and the application of residue integration to examine
only the contribution of the wave traveling in the +$\b{x}$-direction,
we have the transmission amplitudes
  \begin{equation}
    t_{n^\prime,n}^{m^\prime,0} = \delta_{n^\prime,n}\delta_{m^\prime,0}
                                - \frac{i}{2 k_{n^\prime}^{m^\prime}}
                                  T_{n^\prime,n}^{m^\prime,0}(k_{n^\prime}^{m^\prime}, k_n^0)\, .
  \label{tnmm}
  \end{equation}
The two-terminal conductance is simply obtained by the
Landauer-B\"uttiker transmission
function\cite{Landauer1957,Buttiker1982}
\begin{equation}\label{eq:landauer-buttiker}
G(\mu) = \frac{2e^2}{h} \sum_{m^\prime} \mathrm{\mathbf{Tr}} \left[
(\mathbf{t}^{m^\prime})^\dagger \mathbf{t}^{m^\prime} \right] \, ,
\end{equation}
where $\mathbf{t}^{m^\prime}$ is constructed from the transmission
amplitudes (\ref{tnmm}) and our notation indicates transmission matrix contributed
from the sideband $m^\prime$ connecting the incident electron flux in
the various subbands in the source region to the outgoing electron flux
in the subbands in the drain.

\section{Results and discussion}

\begin{figure}[tb]
      \includegraphics[width=0.45\textwidth]{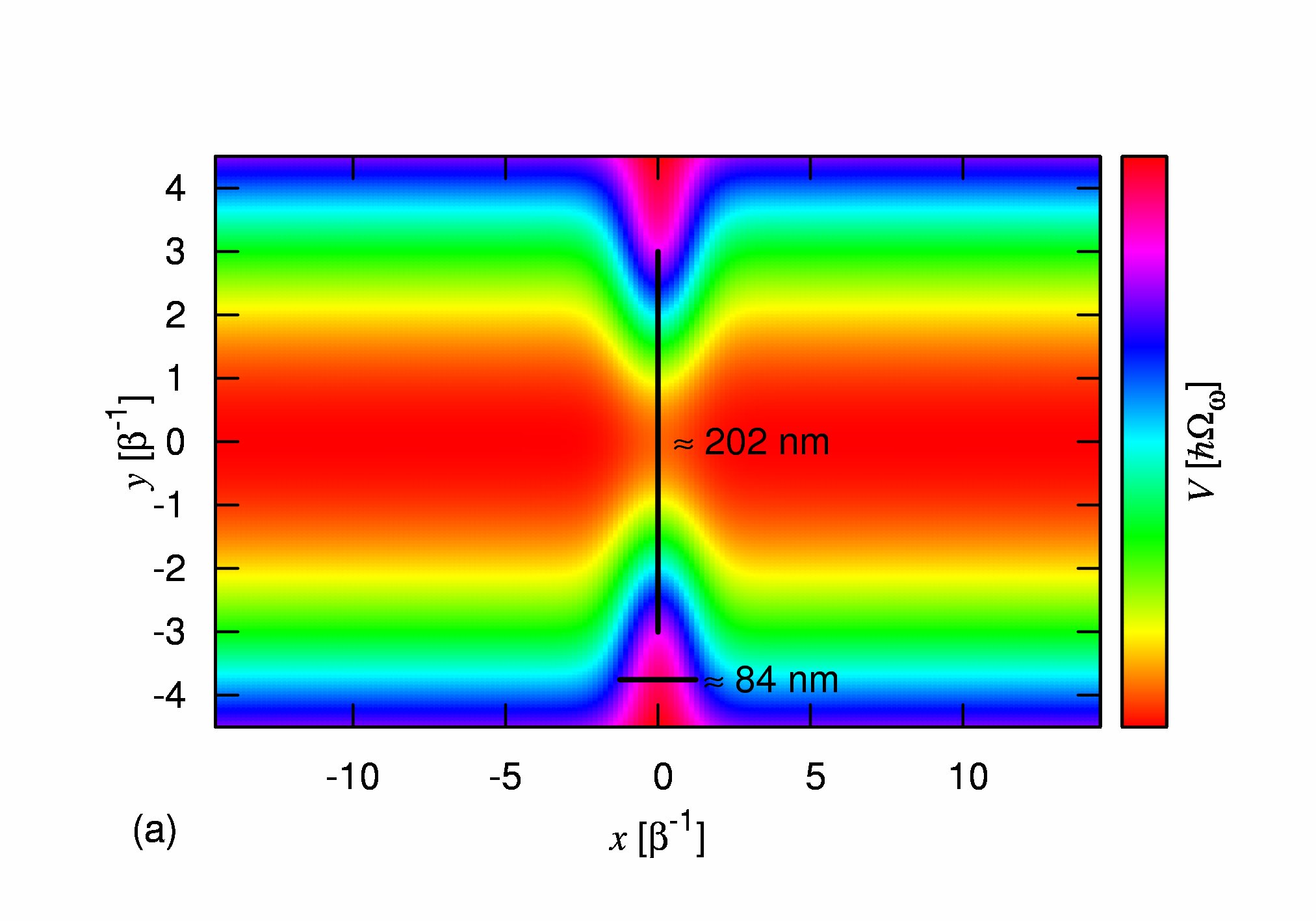}
      \includegraphics[width=0.45\textwidth]{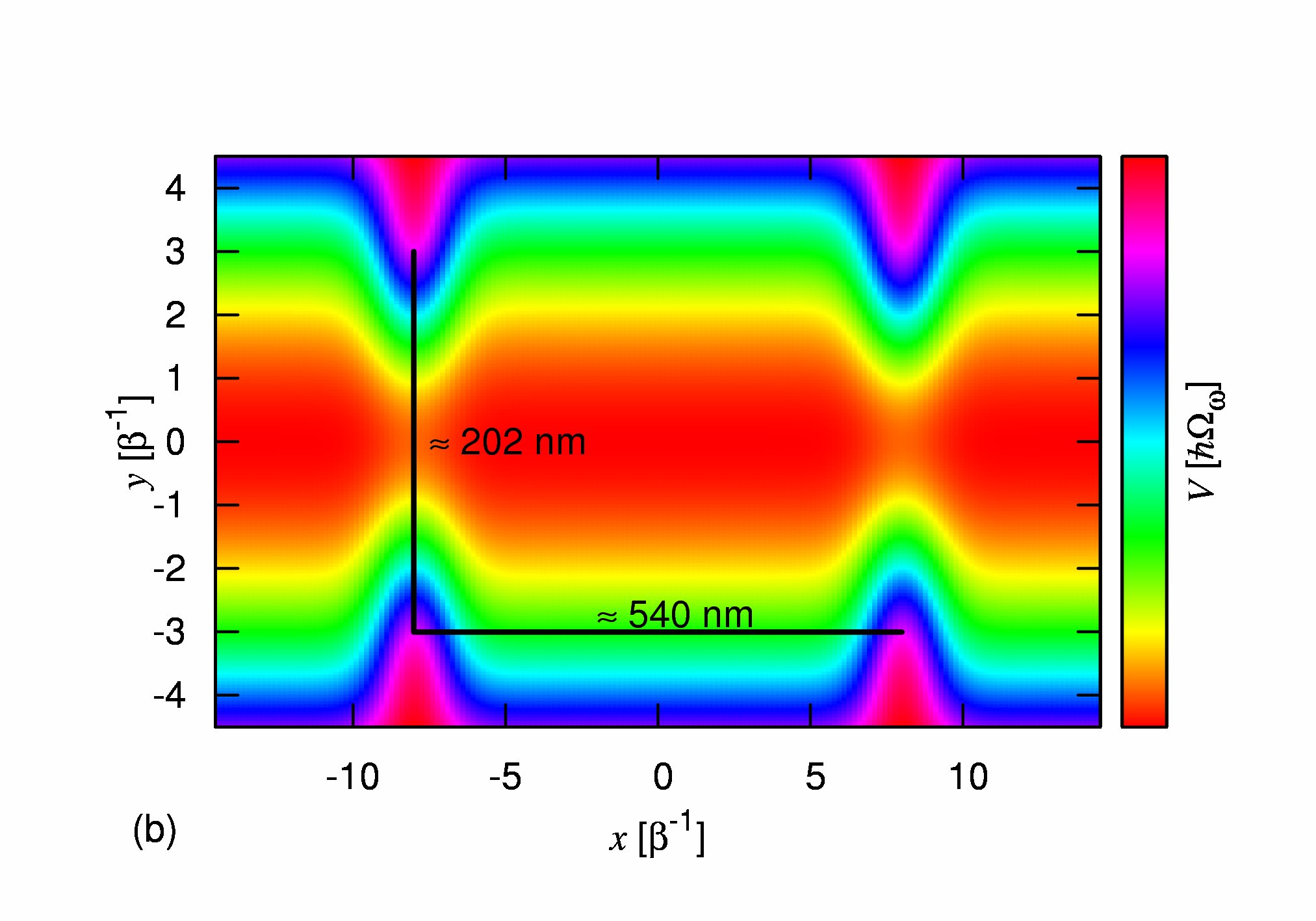}
      \caption{\label{fig:QPC-system}
      (Color online) Schematic diagram of gate-voltage controlled split-gate constriction with no magnetic
      field: (a) Sketch of a single split-gated quantum constriction; (b) Sketch of a double split-gated quantum
      constriction.}
\end{figure}

In this section we investigate magnetotransport properties of
split-gated systems depicted in \fig{fig:QPC-system}.
The considered single QPC (SQPC) system shown in
\fig{fig:QPC-system}(a) can be modeled by the Gaussian-shaped potential
\begin{eqnarray}\label{eq:single-qpc-pot}
  V_{sc}(x,y) &=& V_s e^{-\alpha_x x^2 - \alpha_y (y-y_0)^2}\nn
           &+& V_s e^{-\alpha_x x^2 - \alpha_y (y+y_0)^2}\, ,
\end{eqnarray}
where $y_0$ is the distances of the Gaussian potential peak away from
the center of the wire in the $\b{y}$-direction.  The parameters for
the potential in \eq{eq:single-qpc-pot} are $V_s = 6.5$~meV,
$\alpha_{x} = 0.5\beta_0^2$, $\alpha_{y} = 0.3\beta_0^2$, and $y_0 =
3\beta_0$ such that the width of the QPC is $84$~nm and the distance of
the split-gates is approximately $202$~nm.  The double QPC (DQPC)
system shown in \fig{fig:QPC-system}(b) is described using four
Gaussian-shaped potentials
\begin{eqnarray}\label{eq:double-qpc-pot}
  V_{sc}(x,y) &=& V_s e^{-\alpha_x (x-x_0)^2 - \alpha_y (y-y_0)^2}\nn
              &+& V_s e^{-\alpha_x (x+x_0)^2 - \alpha_y (y+y_0)^2}\nn
              &+& V_s e^{-\alpha_x (x-x_0)^2 - \alpha_y (y+y_0)^2}\nn
              &+& V_s e^{-\alpha_x (x+x_0)^2 - \alpha_y (y-y_0)^2}\, ,
\end{eqnarray}
where $(x_0,y_0)$ are the center coordinates of the Gaussian
potentials.  The parameters are the same as the \eq{eq:single-qpc-pot}
except for $x_0 = 8\beta_0$.

To investigate the electronic transport properties  under a
perpendicular magnetic field, we select the confinement parameter
$\hbar\Omega_0=1$~meV.   We assume that the quantum constriction is
fabricated in a high-mobility GaAs-Al$_x$Ga$_{1-x}$As heterostructure
such that the effective Rydberg energy $E_{\mathrm{Ryd}}=5.92$~meV and
the Bohr radius $a_{\rm B}=9.79$~nm.  Length parameters are scaled
using the the effective magnetic length at zero magnetic field,
referred to as $\beta_0^{-1}$ $(\approx 33.72\,\mathrm{nm})$ while
energy is either fixed in $\mathrm{meV}$ or given in units of the
effective confinement strength $\hbar\Omega_\omega$.

We start by considering a SQPC placed at $x=0$ between two electron
reservoirs, as shown in \fig{fig:QPC-system}(a).  We assume that the
QPC could be induced by metallic split-gates situated on the top of the
heterostructure and can be treated as an open structure with distance
$d_{\rm SG}\approx 202$~nm.
\begin{figure}[tb]
    \includegraphics[width=0.45\textwidth]{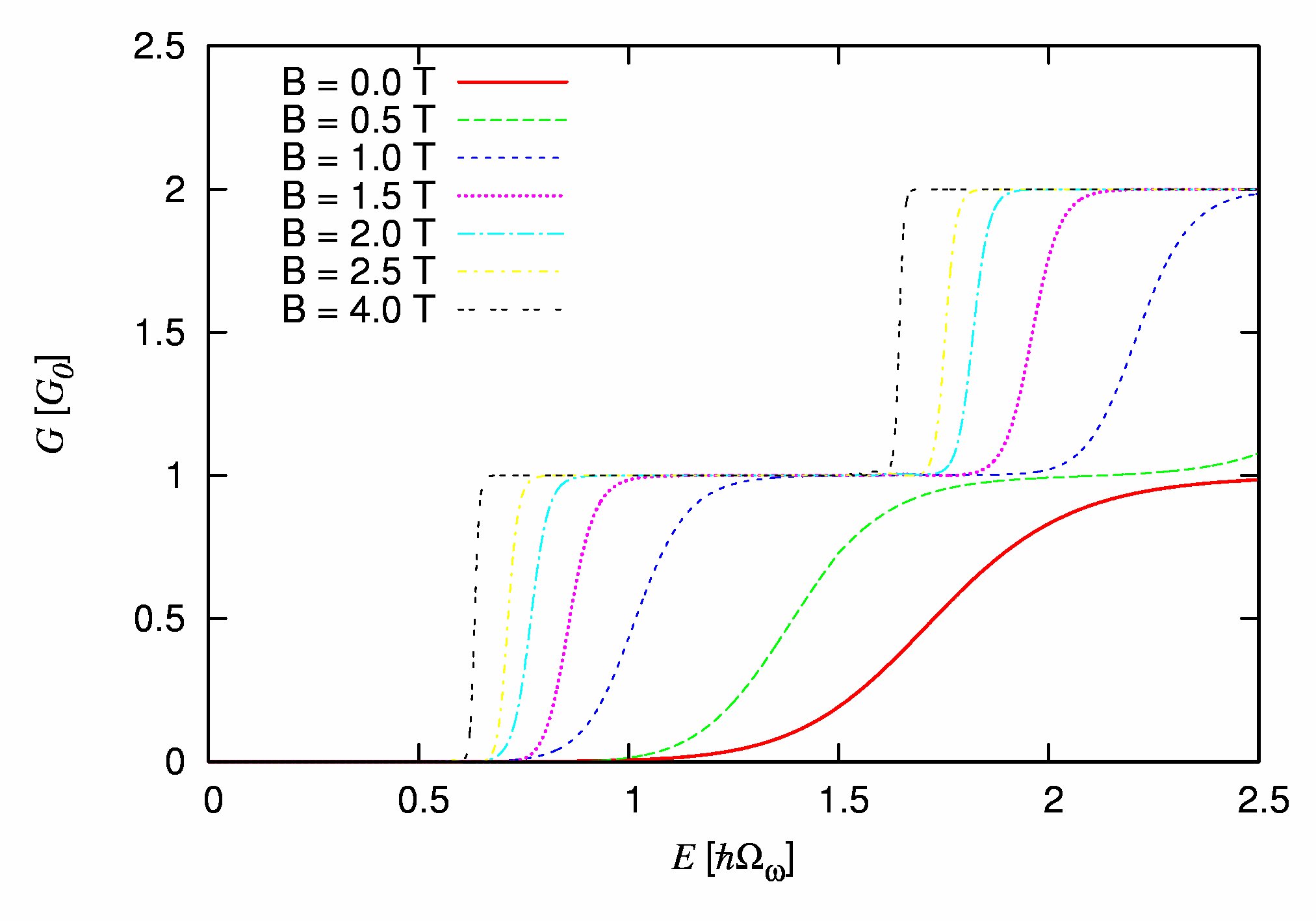}
    \caption{\label{fig:1-QPC-V6.5-Ax0.5-Ay0.3-B0.0-B4.0}
             (Color online) Conductance through a single QPC as a function
             of energy under magnetic field with strength from 0.0 to 4.0
             Tesla. The parameters of the constriction potentials are the
             same with \fig{fig:QPC-system}(a).}
\end{figure}
In \fig{fig:1-QPC-V6.5-Ax0.5-Ay0.3-B0.0-B4.0}, we show the conductance
as a function of incident energy under magnetic field. By increasing
the magnetic field strength from $0.0$ to $4.0$~T, the subband
threshold is red-shifted around $1.1 \hbar\Omega_\omega$, and the
pinch-off regime is also reduced. Moreover, for a given incident
electron energy, increasing the magnetic field may enhance the
conductance which tends to approach the ideal quantization. This is because
of the formation of one-dimensional edge states in the channel
suppressing the backscattering. Since there is no significant
interference, we see that the conductance plateaus are monotonically
increased as a function of energy for arbitrary magnetic fields
implying that no AB oscillations could be induced in such a simple
geometry and small source-drain bias regime.

To enhance the interference effects, we consider a DQPC made by two pairs
of split-gates located at $x = \pm x_0$ with $x_0 = 8\beta_0$ [see
\fig{fig:QPC-system}(b)] forming a cavity with characteristic length
$L\approx 540$~nm.
  \begin{figure}
    \includegraphics[width=0.45\textwidth]{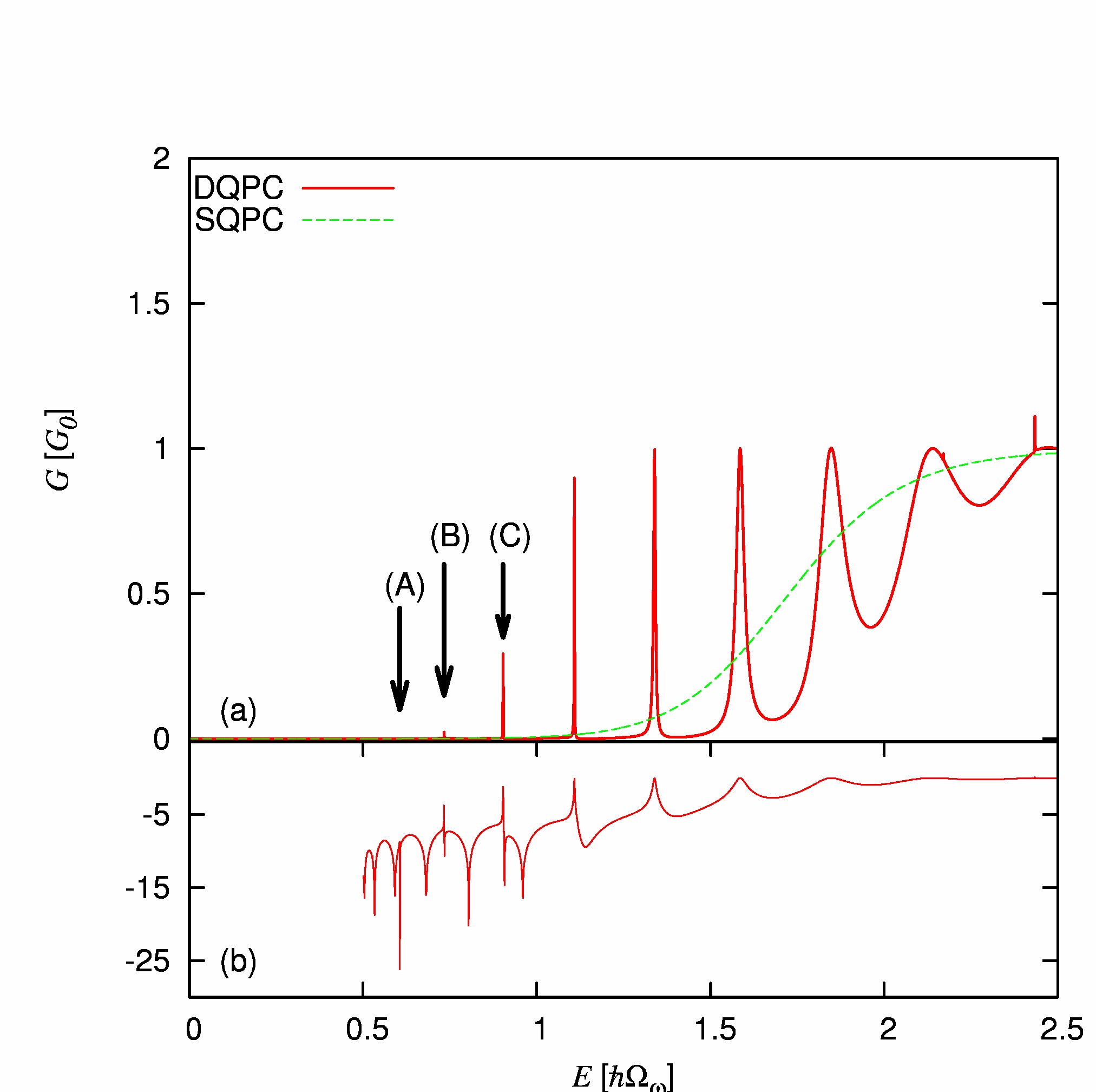}
    \caption{\label{fig:2-QPC-CON-WAVE-V6.5-Ax0.5-Ay0.3-B0.0}
             (Color online) Conductance as a function of energy with no magnetic field:
             (a) The conductance for the case of DQPC (red solid curve) in comparison with the case of SQPC
             (green dashed curve); (b) The logarithm conductance for the case of DQPC.
             The parameters for the potentials of the DQPC are the same as \fig{fig:QPC-system}(b).
             The resonances (A)-(C) shown by  the black arrows are at $E/\hbar\Omega_\omega =
             0.606$, $0.733$, and $0.903$.
             }
  \end{figure}
In \fig{fig:2-QPC-CON-WAVE-V6.5-Ax0.5-Ay0.3-B0.0}(a), we show the
conductance as a function of incident energy in the DQPC system with no
magnetic field (red solid curve) in comparison with the
case of a SQPC (green dashed curve).  By adding the second QPC,
the conductance is strongly suppressed in the non-resonant energies. In
addition, it is interesting to see that the conductance brings forth
resonance peaks instead of dips.\cite{Tang2003} This implies that the
QPC increases the subband threshold, so that electrons with energy in
the pinch-off regime manifest resonant transmission feature induced by
the cavity formed by the DQPC.

\begin{figure}[tb]
    \centering
    \includegraphics[width=0.238\textwidth]{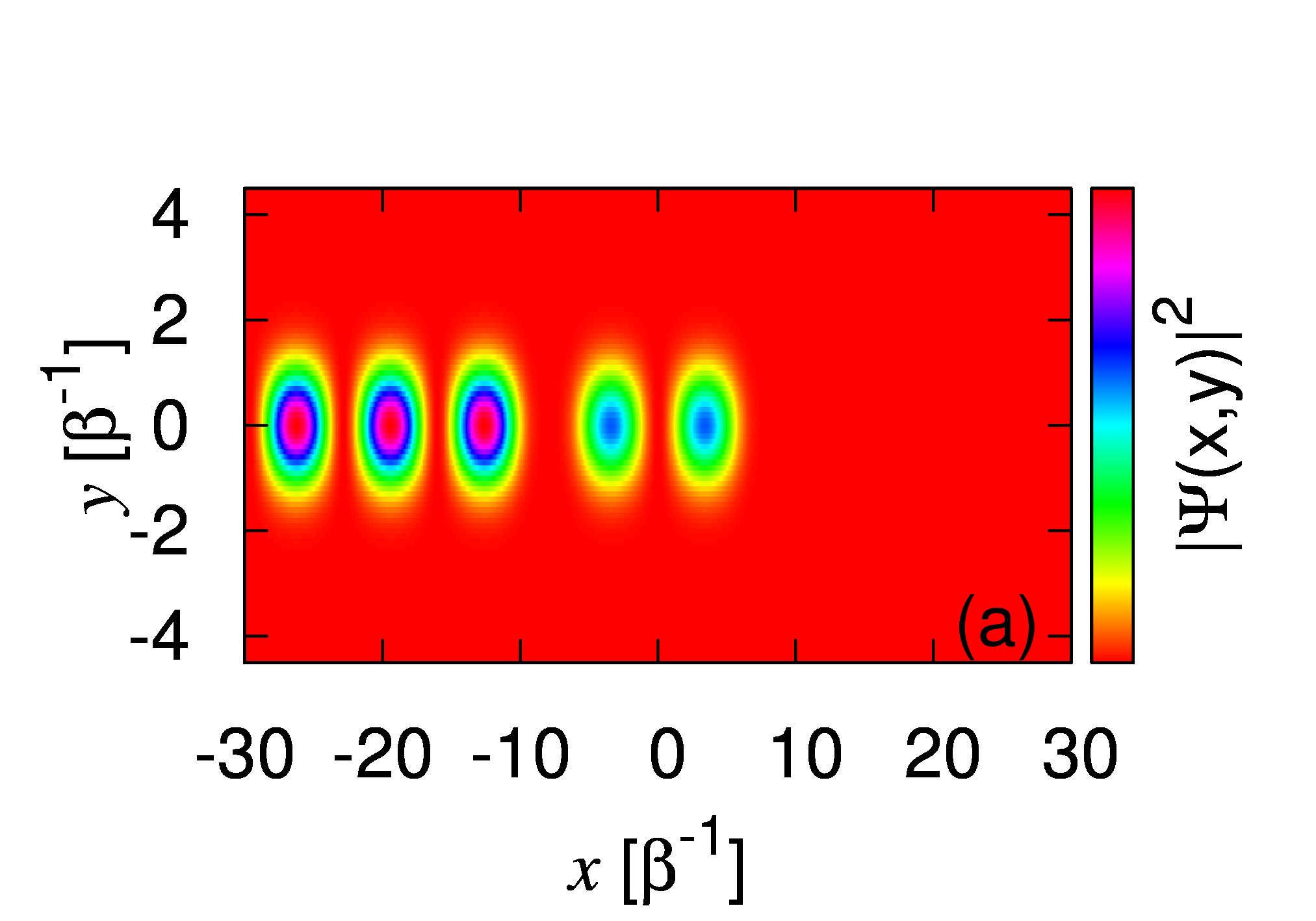}
    \includegraphics[width=0.238\textwidth]{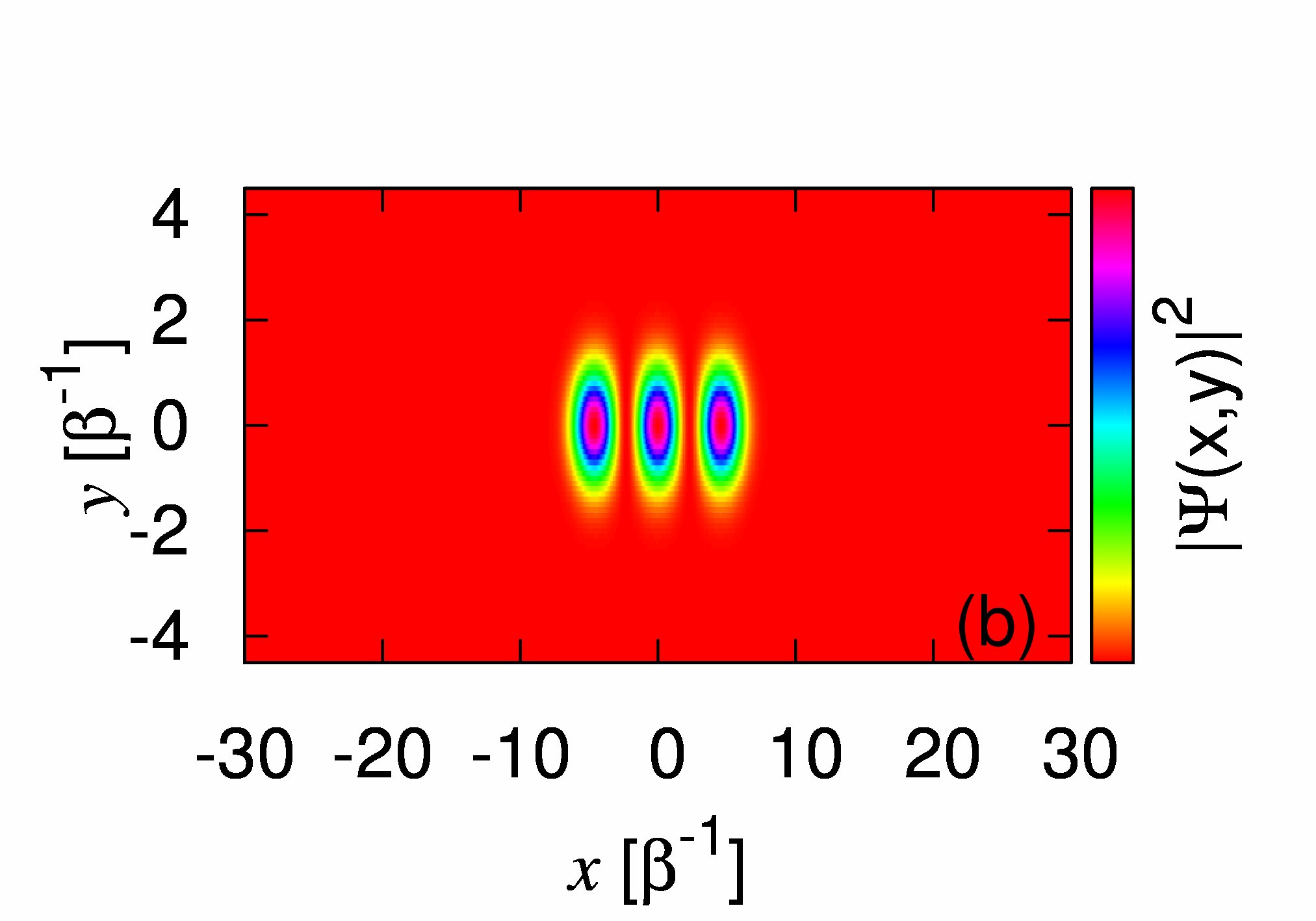}
    \includegraphics[width=0.238\textwidth]{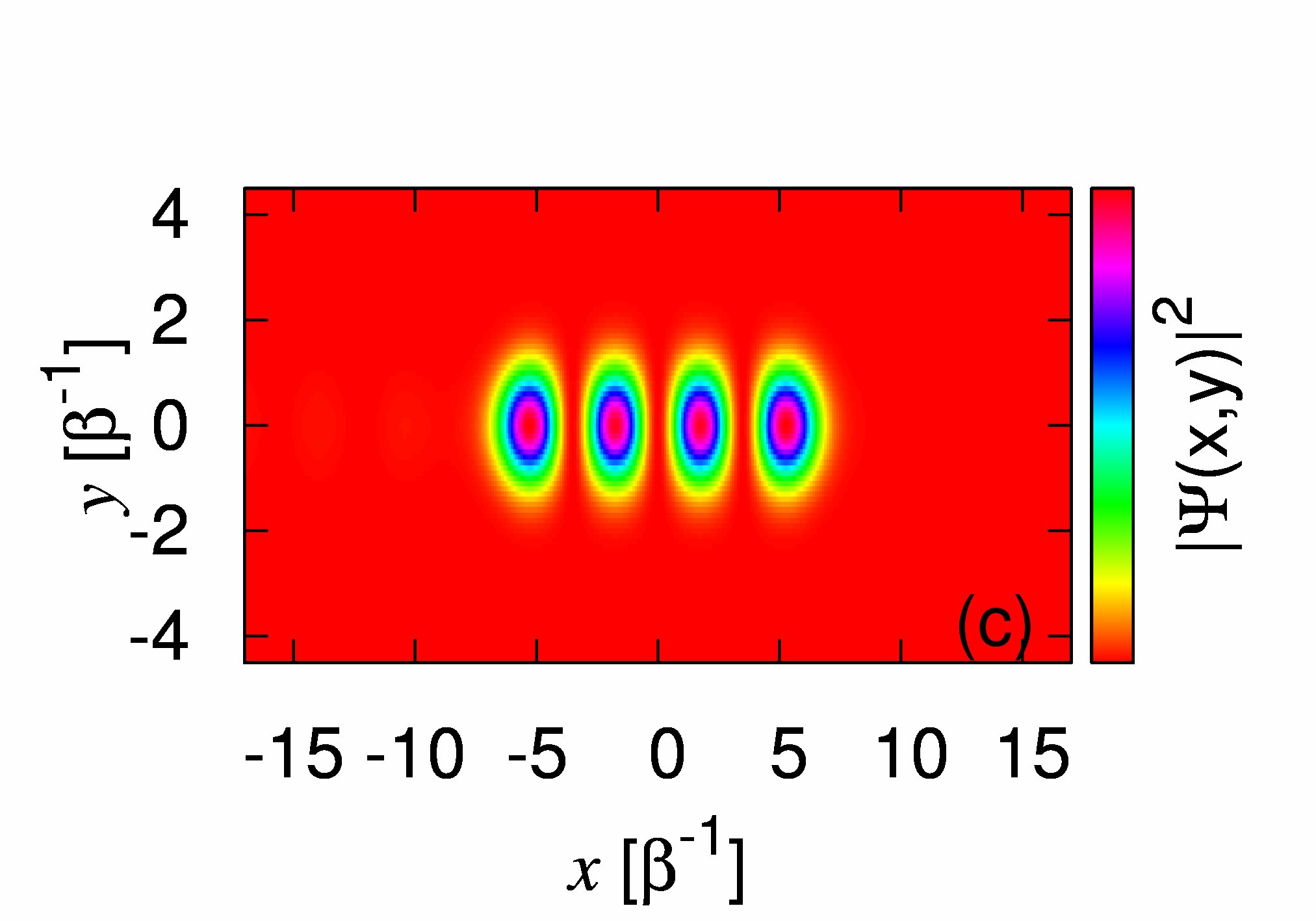}
    \caption{\label{fig:2-QPC-WAVE-V6.5-Ax0.5-Ay0.3-B0.0-QBS}
             (Color online) Probability density for resonances in the conductance marked by (a)-(c) in
             \fig{fig:2-QPC-CON-WAVE-V6.5-Ax0.5-Ay0.3-B0.0}(a)
             with corresponding energies  $E/\hbar\Omega_\omega =$
             (a) $0.606$; (b) $0.733$; (c) $0.903$.}
  \end{figure}

To obtain a deeper understanding for the resonance features shown by the
black arrows in \fig{fig:2-QPC-CON-WAVE-V6.5-Ax0.5-Ay0.3-B0.0}(a), we
plot the logarithm conductance shown in
\fig{fig:2-QPC-CON-WAVE-V6.5-Ax0.5-Ay0.3-B0.0}(b).  It is clearly seen
that the resonance (A) manifests dip structure, while the resonances
(B) and (C) exhibit the Fano line-shapes~\cite{Fano1961} indicating the
interference between the localized and the extended states.  Their
corresponding probability densities are shown in
\fig{fig:2-QPC-WAVE-V6.5-Ax0.5-Ay0.3-B0.0-QBS}(a)-(c).
The number of probability density peaks within the cavity region
implies the order of the resonances formed in the cavity implying that
the resonances (A)-(C) are the second to the fourth resonances in the
cavity.  The ratio of the distance between the nearby peaks to the
incident wave length is around $2.0$, this indicates long-lived
resonance modes fitting the cavity in the DQPC.

We now turn to study the magnetotransport properties in the split-gated
systems.
  \begin{figure}[tb]
    \includegraphics[width=0.45\textwidth]{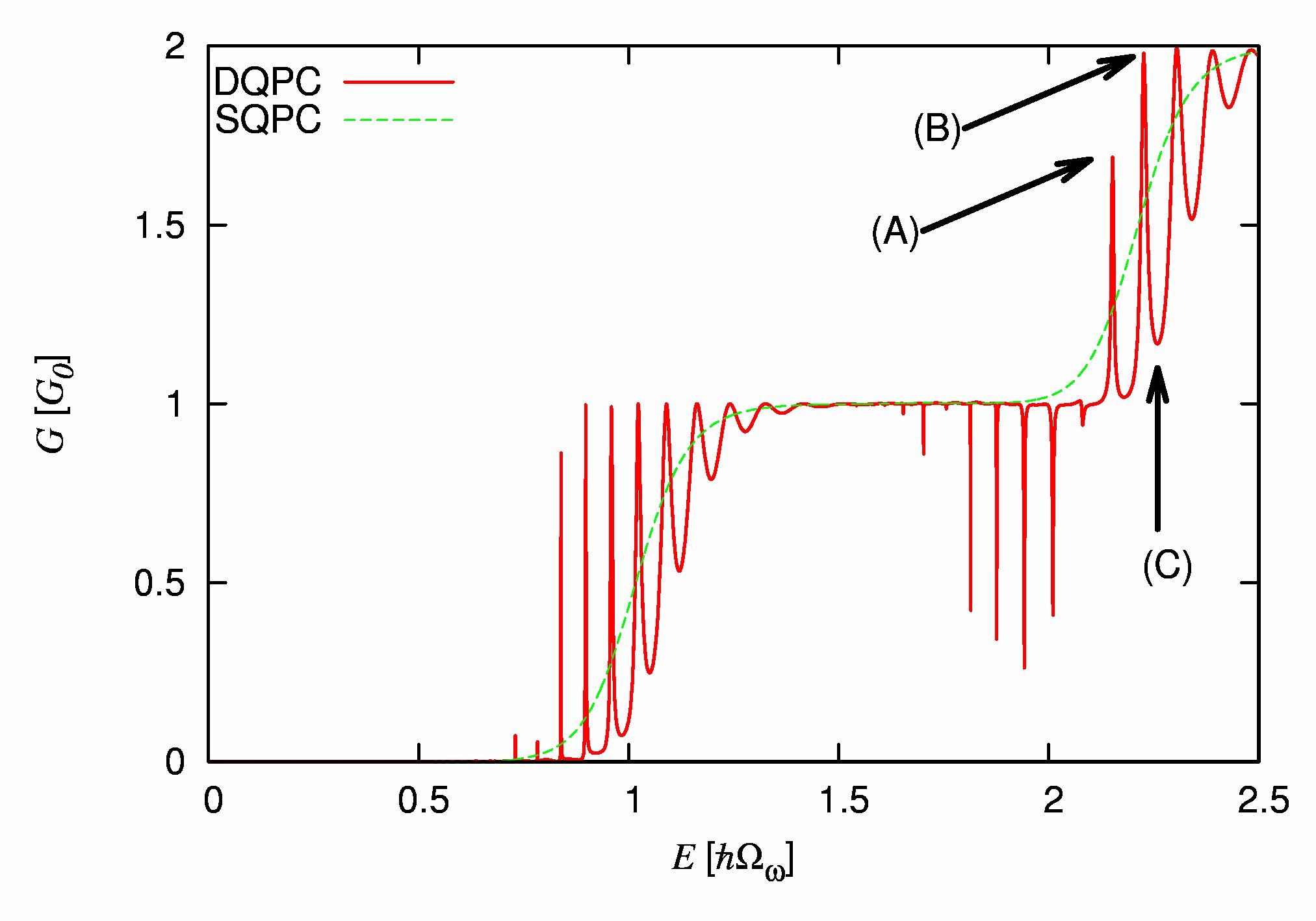}
    \caption{\label{fig:2-QPC-CON-WAVE-V6.5-Ax0.5-Ay0.3-B1.0}
             (Color online) Conductance as a function of energy with $B = 1.0\,\mathrm{T}$.
             The other parameters are the same as previous figures.}
  \end{figure}
In \fig{fig:2-QPC-CON-WAVE-V6.5-Ax0.5-Ay0.3-B1.0}, we present abundant
resonance features in conductance  of SQPC (green dashed curve)
and DQPC (red solid curve) under magnetic field $B = 1.0$ Tesla. For the case of
SQPC, the conductance is monotonically increased in the pinch-off regime
($E/\hbar\Omega_\omega < 1.3$).  The conductance quantization at $1.3 <
E/\hbar\Omega_\omega < 2.0$ demonstrates that electrons can be
transported coherently within the edge channel without significant
backscattering.  For the case of a DQPC, the conductance manifests
resonant transmission peaks in the low kinetic energy regimes of the
first and the second subbands, while the conductance exhibits resonant
reflection features in the high kinetic energy regime. The resonance
structures in the conductance are more dense due to the magnetic field.

\begin{figure}[tb]
    \includegraphics[width=0.238\textwidth]{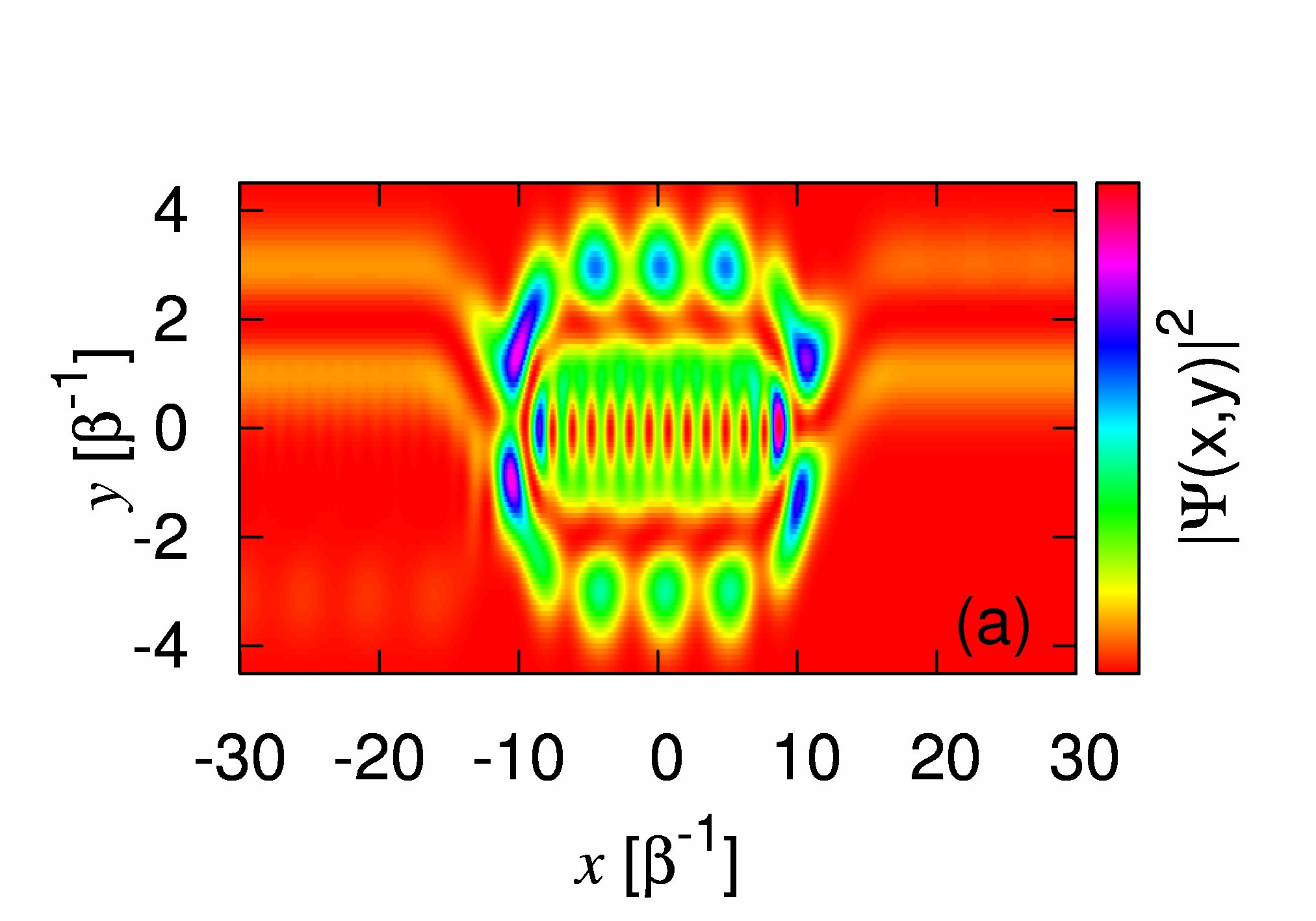}
    \includegraphics[width=0.238\textwidth]{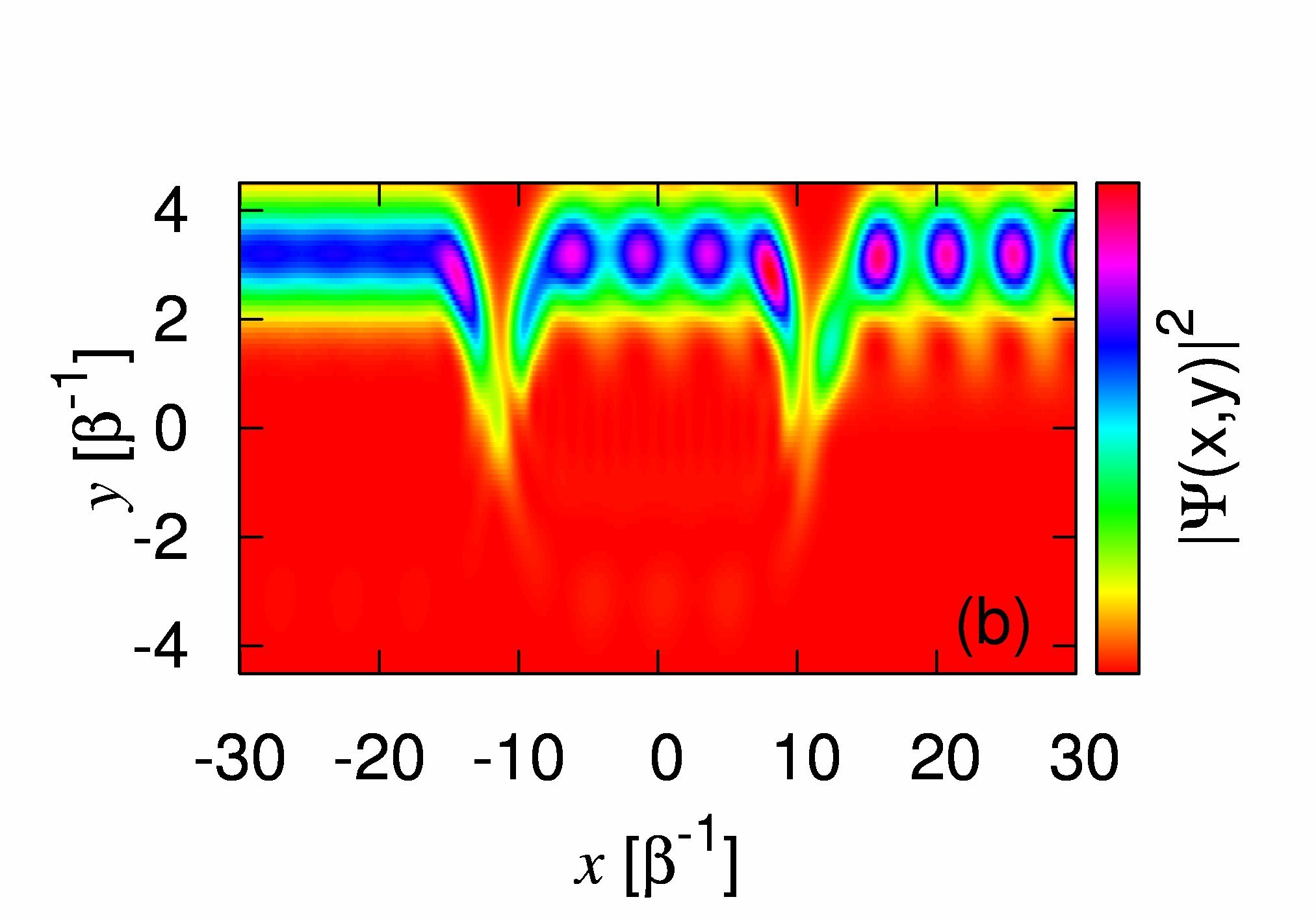}
    \includegraphics[width=0.238\textwidth]{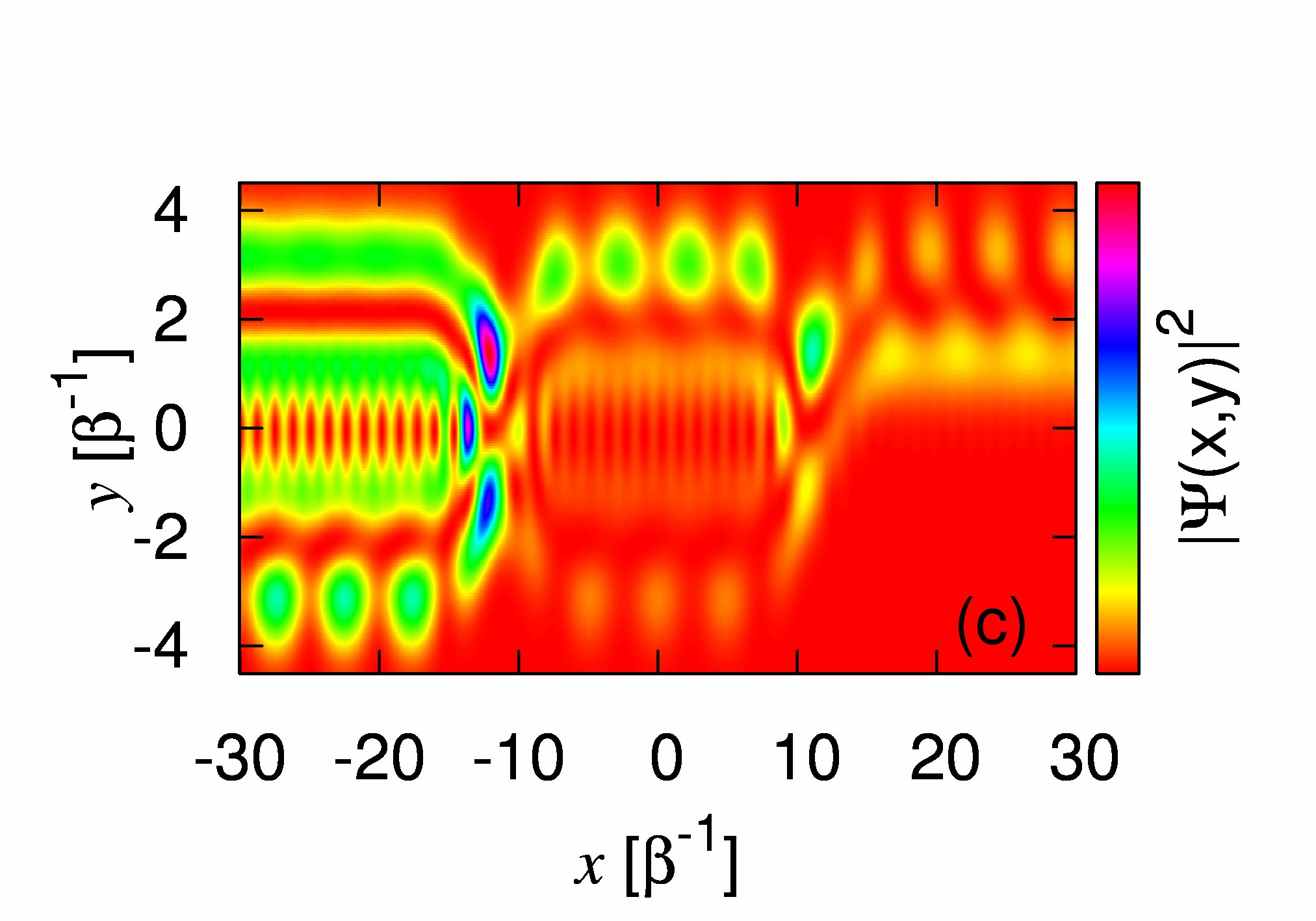}
    \caption{\label{fig:2-QPC-WAVE-V6.5-Ax0.5-Ay0.3-B1.0}
             (Color online) Probability density for the peaks marked in \fig{fig:2-QPC-CON-WAVE-V6.5-Ax0.5-Ay0.3-B1.0}.
             (a) A ring structure inside the cavity due to the magnetic field. $E/\hbar\Omega_\omega = 2.152$, $n = 1$.
             (b) An edge state. $E/\hbar\Omega_\omega = 2.226$, $n = 0$.
             (c) A scattering state. $E/\hbar\Omega_\omega = 2.259$, $n = 1$.}
  \end{figure}

To get a better understanding for the on- resonance peaks (A) and (B) as
well as the off- resonance valley structure (C) marked by the black
arrows in \fig{fig:2-QPC-CON-WAVE-V6.5-Ax0.5-Ay0.3-B1.0}, we plot their
corresponding probability densities in
\fig{fig:2-QPC-WAVE-V6.5-Ax0.5-Ay0.3-B1.0}(a)-(c).
First, when the electron is transported with very low kinetic energy
such as the case of \fig{fig:2-QPC-WAVE-V6.5-Ax0.5-Ay0.3-B1.0}(a)
in which the electron is occupying the second subband $n=1$.  We see
that the localized states in the cavity can be well established forming
double AB-oscillation paths, where the inner path manifests an entangled
feature. In addition, the QBSs can be formed at both ends of the open
cavity. Secondly, if the electron carries sufficient high kinetic
energy such as the case of
\fig{fig:2-QPC-WAVE-V6.5-Ax0.5-Ay0.3-B1.0}(b) in which the
electron is occupying the first subband $n=0$.  The Lorentz force plays
a dominant role on the transport such that the electron wave is pushed to
the upper confinement and forms an edge state facilitating the flow of
electrons through the system by suppressing backscattering in the
system. For comparison with the case (A), we show the non-resonance
probability feature shown in
\fig{fig:2-QPC-WAVE-V6.5-Ax0.5-Ay0.3-B1.0}(c) in which the electron is
also occupying the second subband $n=1$.  It is important that, in the
non-resonant condition, the Lorentz force is able to push the electron
a little bit to the upper confinement and QBS can be formed only at
the left QPC thus manifesting reflection feature with minimal
conductance.

  \begin{figure}[t]
    \includegraphics[width=0.4\textwidth]{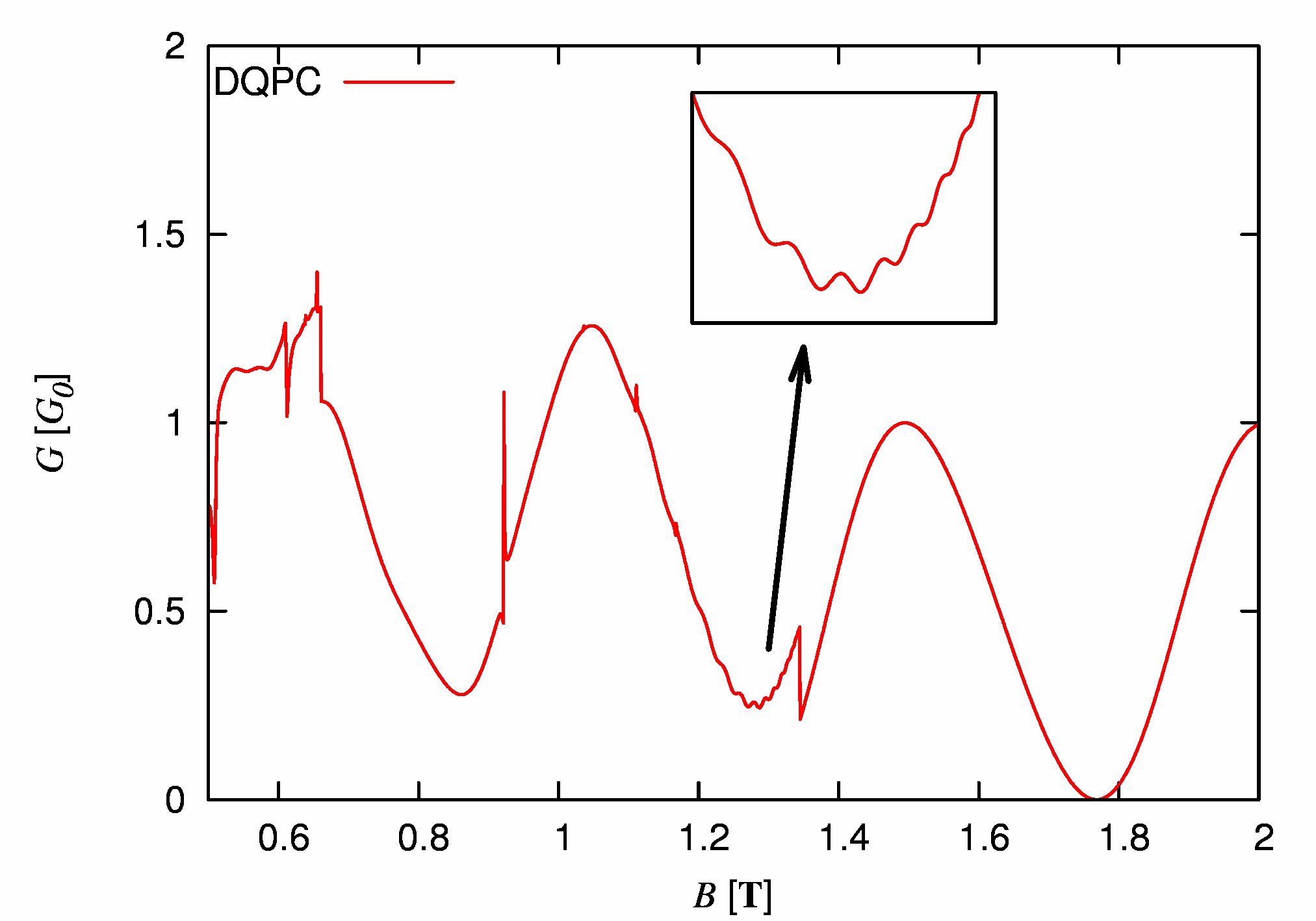}
    \caption{\label{fig:2-QPC-CON-AB-B1.0-E1.9}
             (Color online) Conductance $G$ as a function of magnetic field $B$ for electrons with
             incident energy $E/\hbar\Omega_\omega = 1.9$ through DQPC.  The inset shows that the
             conductance with small AB oscillations can be seen superimposed on large oscillations.
             The other parameters are $V_s = 6.5\,\mathrm{meV}$, $\alpha_{x} = 0.5\beta_0^2$, $\alpha_{y} = 0.3\beta_0^2$,
             $y_0 = 3\beta_0$, and $x_0 = 8\beta_0$.}
  \end{figure}

\begin{figure}[b]
    \includegraphics[width=0.3\textwidth]{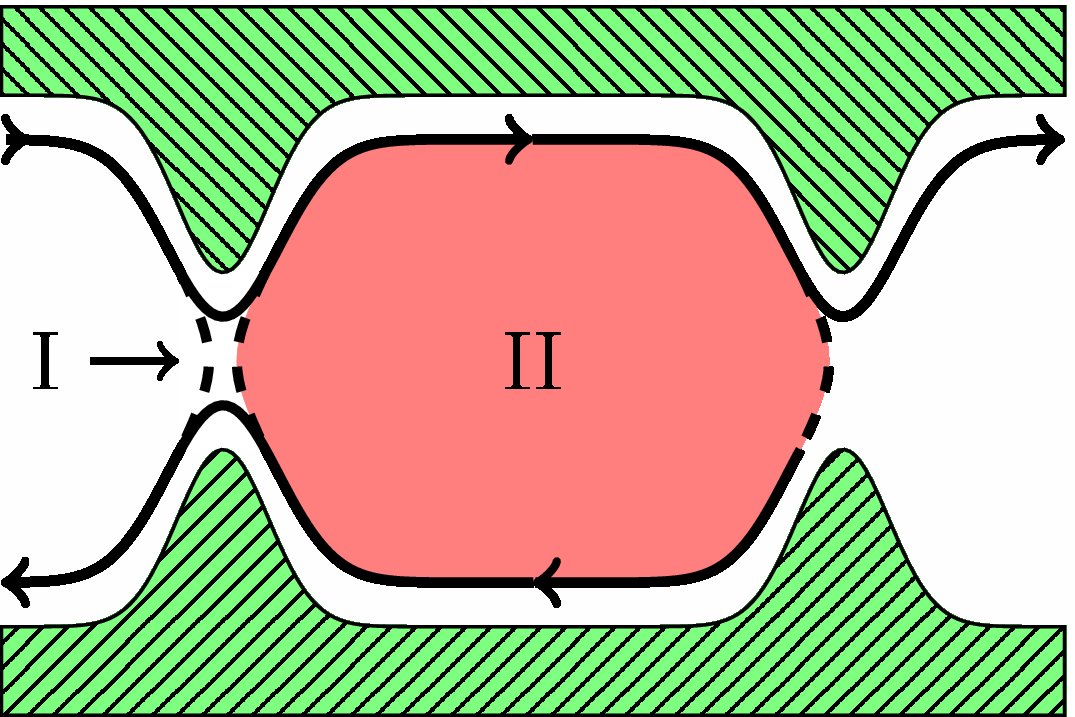}
    \caption{\label{fig:ab-effect}
             (Color online) Illustration of the possible paths that the
             electrons can take and the paths that they enclose in the
             DQPC system confined by double split gates (shaded regions).}
  \end{figure}

In \fig{fig:2-QPC-CON-AB-B1.0-E1.9}, we show that the conductance
versus magnetic field exhibits periodic oscillations.  The period
$\Delta B$ of AB oscillations is inversely proportional to the
effective area ${\cal A}$ enclosed by the electron path, given by
$\Delta B = \Phi_0{\cal A}^{-1}$ with $\Phi_0 = h/e$ being the flux
quantum.~\cite{Ihnatsenka2008} The effective area can be slightly
changed by tuning the strength $V_s$ of split gates.  The AB
oscillations with large period $\Delta B \approx 0.5$ Tesla is
associated with the interference between the directly reflected
electrons by the left QPC and the electrons go through an enclosed path
forming a small area in the left QPC (area I in \fig{fig:ab-effect}).
The small oscillations superimposed on
the larger ones shown in the inset of \fig{fig:2-QPC-CON-AB-B1.0-E1.9}
are formed due to the interference between the electrons directly
reflected by the right QPC and the electrons going through the open cavity
forming a large area in the DQPC (area II in \fig{fig:ab-effect}).
We note in passing that our results demonstrate that the AB
oscillations do not require a ring geometry.~\cite{Loosdrecht88:10162}
Interference is the most important effect to generate AB oscillations.
Moreover, the robust zero conductance feature at around $B\simeq
1.77$~Tesla exhibits that the DQPC could be applicable as a mesoscopic
switching device.

To explore the time-dependent transport in a DQPC with time-harmonic
modulation, we construct the model by using four Gaussian-shaped
potentials that are expressed as
\begin{eqnarray}
  V_{sc}(x,y,t) &=& V_{\rm L}(t) e^{-\alpha_x (x-x_0)^2 - \alpha_y (y-y_0)^2} \nn
                &+& V_{\rm L}(t) e^{-\alpha_x (x+x_0)^2 - \alpha_y (y+y_0)^2} \nn
                &+& V_{\rm R}(t) e^{-\alpha_x (x-x_0)^2 - \alpha_y (y+y_0)^2} \nn
                &+& V_{\rm R}(t) e^{-\alpha_x (x+x_0)^2 - \alpha_y (y-y_0)^2}\, ,
\end{eqnarray}
where strengths of the left QPC $V_{\rm L}(t) = V_s + V_t \cos{\omega
t}$ and the right QPC $V_{\rm R}(t) = V_s + V_t \cos{(\omega t +
\phi_0)}$ contain the same driving frequency $\omega$ with a phase
difference $\phi_0$. This driven DQPC system is similar to the one
depicted in \fig{fig:QPC-system}(b) except for the external
driving terms with amplitude $V_t$.
  \begin{figure}[tb]
    \includegraphics[width=0.45\textwidth]{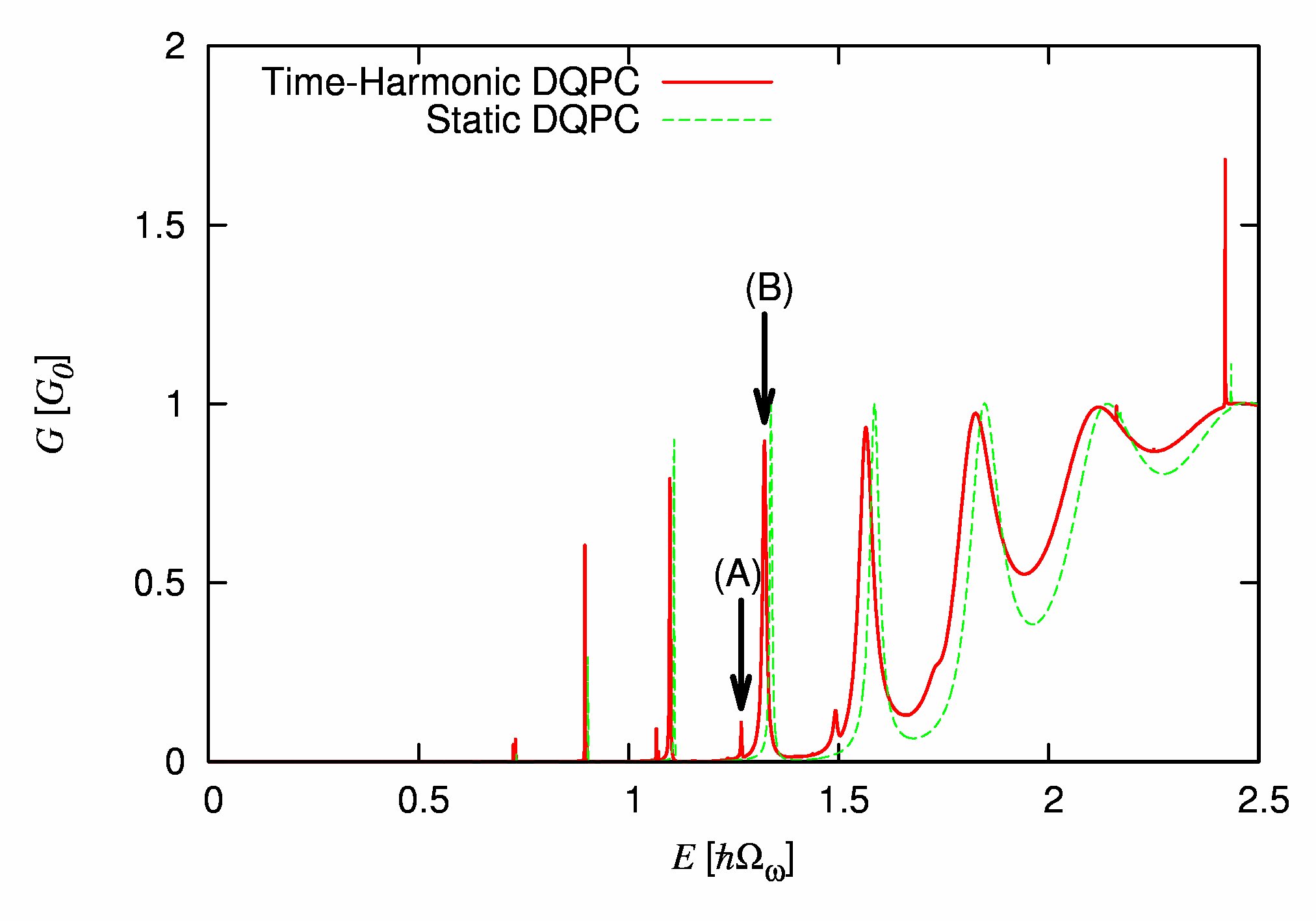}
    \caption{\label{fig:2-QPC-D-Vs6.0Vt0.5-B0.0-PHIpi}
             (Color online) Conductance versus incident energy with no magnetic field ($B=0$)
             for a time-harmonic DQPC ($V_t = 0.5$~meV and $\phi_0 = \pi$, red solid)
             in comparison with that of a static DQPC ($V_t = 0.0$~meV, green dashed).
             The other parameters for the system are $V_s = 6.0\,\mathrm{meV}$,
             $\alpha_{x} = 0.5\beta_0^2$, $\alpha_{y} = 0.3\beta_0^2$, $y_0 = 3\beta_0$, $x_0 = 8\beta_0$,
             and driving frequency $\omega = 0.17 \Omega_\omega$.}
  \end{figure}
  \begin{figure}[tb]
    \includegraphics[width=0.45\textwidth]{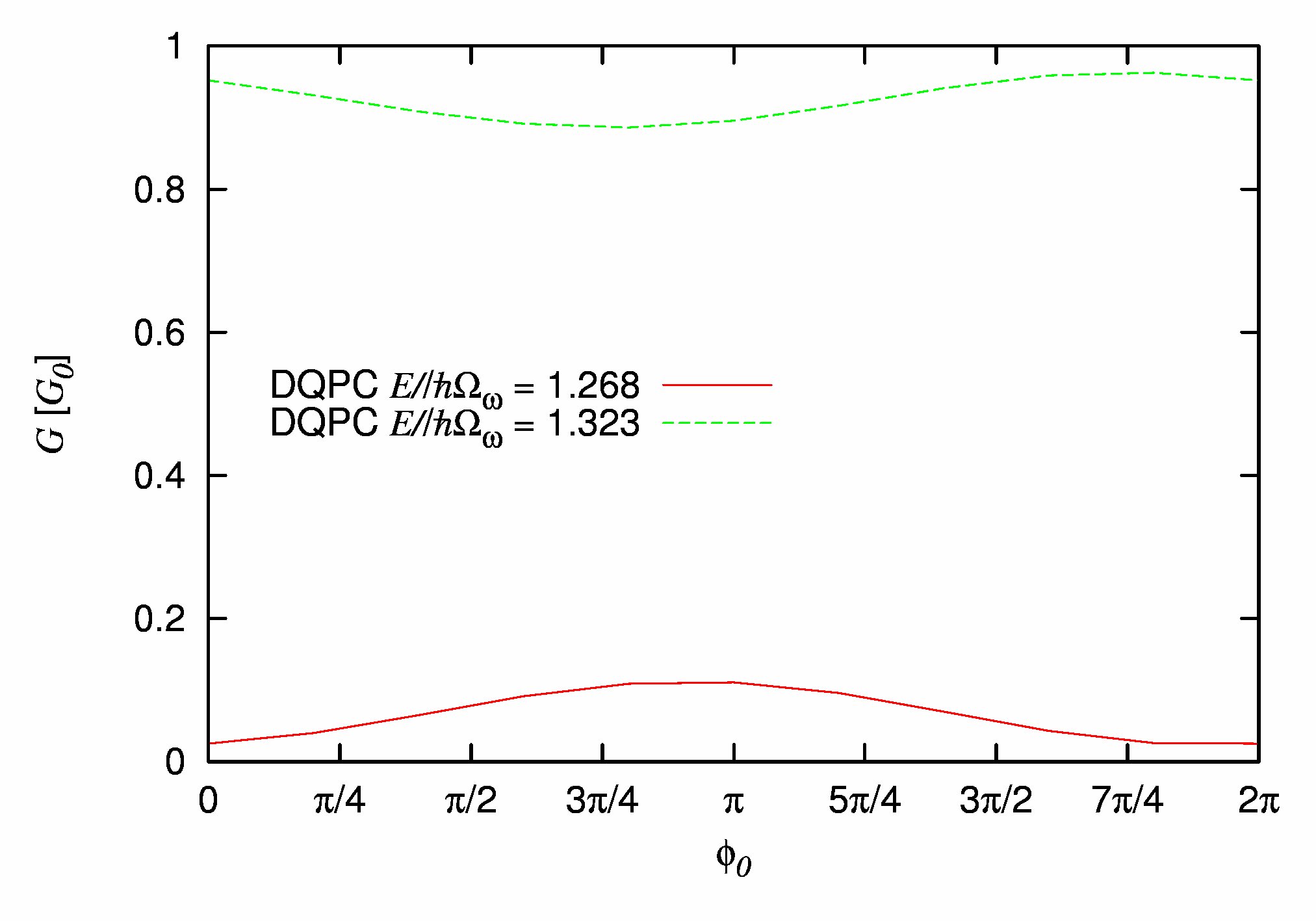}
    \caption{\label{fig:2-QPC-P-Vs6.0Vt0.5-B0.0}
             Conductance $G$ versus phase difference $\phi_0$ between the left and the right QPCs
             for the electron with energies marked by (A) and (B) in
             \fig{fig:2-QPC-D-Vs6.0Vt0.5-B0.0-PHIpi}.}
  \end{figure}

In \fig{fig:2-QPC-D-Vs6.0Vt0.5-B0.0-PHIpi}, we show the conductance as
a function of incident energy for the time-harmonic DQPC ($V_t =
0.5$~meV and $\phi_0 = \pi$, red solid curve) in comparison with the static
DQPC ($V_t = 0.0$~meV, green dashed curve).  Under the influence of the
time-harmonic driving potential, we find a small side peak in
$G$ marked by (A) indicating that the electron is allowed to emit a
photon with energy $\hbar\omega$ and jump to a state beneath the resonance, i.e.,
the main peak marked by (B).  Moreover, the electron kinetic energy
plays a role to suppress such inter-sideband transitions. As we can see
that the small peak becomes a shoulder structure for an electron with
incident energy at around $E \simeq 1.75 \hbar\Omega_\omega$.
In \fig{fig:2-QPC-P-Vs6.0Vt0.5-B0.0},  we show
the conductance as a function of phase difference between the two QPCs.
The energies are fixed at the main peak $E=1.323 \hbar\Omega_\omega$
(green dashed curve) and at the side peak $E=1.268
\hbar\Omega_\omega$ (red solid curve) marked, respectively,
by (A) and (B) in \fig{fig:2-QPC-D-Vs6.0Vt0.5-B0.0-PHIpi}.
Both cases are not very sensitive to the phase difference $\phi_0$, but we see that
$\phi_0$ at around $\pi$ can enhance the inter-sideband transitions.

\section{Concluding Remarks}

We have developed a Lippmann-Schwinger model that has allowed us to
explore the magnetotransport and time-dependent transport spectroscopy
of coherent elastic and inelastic multiple scattering features
relevant to quantum constricted SQPC and DQPC systems under a magnetic
field perpendicular to the 2DEG.  We have demonstrated and analyzed the
mechanisms causing the slow and the fast conductance oscillations due
to AB interference in the DQPC system. We hope that our numerical
demonstrations on magnetotransport and time-dependent transport could
be useful for the utilization of intricate coupling between
subbands and sidebands towards the realization of quantum pumping
circuits and fast manipulation of quantum information processing in
mesoscopic systems.

\begin{acknowledgments}
This work was financially supported by the Research and Instruments
Funds of the Icelandic State,  the Research Fund of the University of
Iceland, the  Icelandic Science and Technology Research Programme for
Postgenomic Biomedicine, Nanoscience and Nanotechnology, and the
National Science Council of the Republic of China through Contract No.
NSC97-2112-M-239-003-MY3.
\end{acknowledgments}

\end{document}